# Coupled orbital angular momentum conversions in a quasi-periodically poled LiTaO$_3$ crystal

Xinyuan Fang[1], Guang Yang[1], Dunzhao Wei[1], Dan Wei[1], Rui Ni[1], Wei Ji[1], Xiaopeng Hu[1], Wei Hu[1], Yong Zhang[1,a)], S. N. Zhu[1], and Min Xiao[1,2,b)]

[1]National Laboratory of Solid State Microstructures, College of Engineering and Applied Sciences and School of Physics, Nanjing University, Nanjing 210093, China
[2]Department of Physics, University of Arkansas, Fayetteville, Arkansas 72701, USA
* Corresponding author: a) Electronic mail: zhangyong@nju.edu.cn
b) Electronic mail: mxiao@uark.edu

**We experimentally demonstrate the orbital angular momentum (OAM) conversion by the coupled nonlinear optical processes in a quasi-periodically poled LiTaO$_3$ crystal. In such crystal, third-harmonic generation (THG) is realized by the coupled second-harmonic generation (SHG) and sum-frequency generation (SFG) processes, i.e., SHG is dependent on SFG and vice versa. The OAMs of the interacting waves are proved to be conserved in such coupled nonlinear optical processes. As increasing the input OAM in the experiment, the conversion efficiency decreases because of the reduced fundamental power intensity. Our results provide better understanding for the OAM conversions, which can be used to efficiently produce an optical OAM state at a short wavelength.**

In 1992, L. Allen pointed out that the light beam with an azimuthal phase dependence of exp(il$\varphi$) carries an orbital angular momentum (OAM), where $l$ is the azimuthal mode index [1]. Such beam can be experimentally produced by various ways such as Q-plate (QP) [2], spiral phase plates [3], holographic diffraction gratings [4], and segmented adaptive mirrors [5]. Significant attentions have been focused on OAM because it can be widely used in optical tweezers [6,7], optical manipulation [8], optical trapping [9], imaging [10] and information processing [11-15]. Recently, OAM beams are applied in optical communication to increase the channel capacity and the spectral efficiency [16]. In practical applications, one often need imprint OAM onto a light beam with a short wavelength such as in UV band, which is, however, inconvenient to be realized through most of the traditional methods described above. Nonlinear optical conversion of an OAM state is an alternative and feasible way. Experimental demonstrations have been achieved in second-harmonic generation (SHG), sum-frequency generation (SFG), high-harmonic generation (HHG) [17], and spontaneous parametric down conversion (SPDC) [18-20]. It is an important issue to understand how the OAM evolves during nonlinear optical conversions. In 1996, Padgett et al. reported the conservation law of OAM in a SHG process through the birefringent phase match (BPM) method [21,22]. Since then, the OAM conservation has been proved in most nonlinear optical interactions with a few exceptions in the SPDC processes.

Recently, periodically-poled LiTaO$_3$ (PPLT) crystals are used to realize OAM conversions [23,24]. PPLT crystals have been widely studied in the past decades because they can efficiently realize frequency conversion through the quasi-phase-matching (QPM) technique [25]. Numerous interesting phenomena have been discovered in 1D and 2D PPLT crystals. [26-28] Comparing to the BPM method, QPM can greatly release the phase-matching requirement by introducing reciprocal vectors. It can also utilize the largest nonlinear optical coefficients, for example, d$_{33}$ in the LiTaO$_3$ crystal. The theory of QPM OAM conversion is proposed by Shao et al. with the help of the coupling-wave equations [29]. The experiments have been carried out through QPM SHG and SFG processes [23,30], which agree well with the OAM conservation. Interestingly, one can realize multiple copies of second-harmonic (SH) OAM states in a 2D PPLT crystal, which presents that the OAM conservation has certain tolerance for phase mismatch between the interacting waves. However, it still remains undiscovered how an OAM state develops in coupled nonlinear optical conversions. In this letter, we investigate the coupled OAM conversions through a QPM third harmonic generation (THG) process in a quasi-periodically poled LiTaO$_3$ (QPPLT) crystal.

An efficient THG [31-33] can be achieved by cascading a SHG process and a SFG process. Usually, two PPLT crystals are required to compensate the phase mismatches in SHG and the cascaded SFG, respectively [32]. However, in a single QPPLT crystal the THG can be realized by coupling the SHG and SFG processes. The QPPLT crystal in our experiment is shown in Fig. 1a. It consists of two fundamental blocks A and B arranged according to the Fibonacci sequence, i.e., ABAAB…, as shown in Fig. 1b. As shown in Fig. 1b, $l_A$ and $l_B$ are the widths of A and B, respectively. Every block contains a pair of antiparallel 180° domains, $l_A = l_{A1} + l_{A2}$ and $l_B = l_{B1} + l_{B2}$. $l_{A1} = l_{B1} = l$ for the positive domain of the sample while $l_{A2} = l(1+\eta)$ and

$l_{B2} = l(1-\tau\eta)$ in the negative domain. The parameters $l$, $\eta$ and $\tau$ are set according to our requirements. The average structure parameter D is defined to be $\tau l_A + l_B$. The reciprocal vector in such structure is defined by $G_{m,n} = 2\pi D^{-1}(m+n\tau)$ with integral m and n. The QPPLT structure can simultaneously fulfill the QPM THG condition which includes SHG and SFG processes (Fig. 1c). Although the QPM configuration is similar to the THG case using two separate PPLT structures [32], the generations of SH and TH waves in a QPPLT crystal are coupled with each other, i.e., SH wave is dependent on the TH wave and vice versa (see the coupling wave equations in [34]). It is interesting to investigate the OAM conversion in such system.

The experimental setup is depicted in Fig. 2. The input fundamental field is generated by an optical parametric oscillator (Horizon I-8572, Continuum Co.) pumped by a nanosecond laser system with a pulse width of about 6 ns and a repetition rate of 10 HZ. The input wavelength is set at 1582 nm. To realize THG, the structure of the QPPLT crystal is designed to be $l_A$ = 21.69 $\mu m$, $l_B$ = 15.29 $\mu m$, and $\tau$ = 5.076. The phase-matching is achieved by involving $G_{1,1}$ for the SHG process and $G_{2,3}$ in the process of SFG (Fig. 1c). In the experiment, we imprint OAM on the input fundamental beam with a QP. The QP used here is a half-wave plate fabricated by a birefringent liquid crystal with a space-variant optical axis in the transverse plane [2]. The geometry of the optical axis is defined by a topological charge "q" which is an integer or a semi-integer. When a circularly-polarized light beam passes such a QP, an OAM of 2q is transferred into the beam. In our experimental setup, the first quarter-wave plate (QWP) is used to change the linear polarization of the input laser to a circular polarization. After planting the OAM information through the QP, another QWP transforms the polarization of the generated vortex beam back to a linear polarization along the z-axis (Fig. 2). Then, the fundamental wave with a known topological charge is focused on the QPPLT slice. Under the configuration, the involved nonlinear optical coefficient is $d_{33}$, which is modulated in the QPPLT crystal. The obtained SH and TH patterns are collected by a CCD camera after filtering out the fundamental beam. A cylindrical lens is used as a mode converter to analyze the OAM information from the QPM THG pattern. By counting the dark stripes in the converted pattern, one can obtain the topological charge of the OAM states [35].

The experimental images recorded on the CCD camera are shown in Fig. 3. Firstly, the fundamental beam is imprinted with an OAM of $l_1$ = 1. The observed SH and TH patterns are shown in Figs. 3(a) and 3(b), which are converted by using a cylindrical lens as shown in Figs. 3(c) and 3(d), respectively. By counting the dark stripes in the converted patterns, the topological charges of the generated SH and TH beams are $l_2$ = 2 and $l_3$ = 3, respectively. Obviously, the OAM conversion in the QPM THG process, i.e., the coupled SHG and SFG processes, follows the conservation law

$$l_2 = 2l_1$$
$$l_3 = 3l_1 \quad (1)$$

To further prove the OAM conservation in such coupled nonlinear optical interactions, we change the input OAM to be $l_1$ = 2 and $l_1$ = 3, which can produce TH beams with $l_3$ = 6 (Fig. 4a) and $l_3$ = 9 (Fig. 4b), respectively. Our experimental results indicate that the coupling between different nonlinear optical processes in a $\chi^{(2)}$ modulated crystal does not change the OAM conservation.

The conversion efficiency of the input beam carrying different OAM is shown in Fig. 5a. Obviously, the maximum conversion efficiencies are 33.3% for SHG and 8.2% for THG, which are achieved when no OAMs are imprinted. When the topological charge of the input OAM state increases, the conversion efficiency of SHG or THG clearly decreases. As show in Fig. 5a, 1.2%, 1%, and 0.5% conversion efficiency for THG and 24.5%, 22.5%, 19.8% conversion efficiency for SHG are obtained in our experiment, which are corresponding to an input $l_1$ = 1, 2, and 3, respectively. This can be explained by the change in the fundamental power density after the introduction of OAM. In our experiment, an input OAM state with a higher topological charge has a ring-shaped intensity distribution with a bigger diameter, which results in a lower fundamental power density. Figure 5b shows the temperature tuning curves of the SHG and THG processes. The peak intensities of the SH and TH beams in our experiment can be achieved at 126 degrees centigrade and 133 degrees centigrade, respectively. The difference in the temperature may originating from the non-perfect dispersion law used to design the QPPLT structure, which indicates that the QPM conditions for SHG and SFG in Fig. 1c cannot be totally fulfilled at the same time. As a result, the OAM conversion is less efficient in the experiment. The conversion efficiency can be further improved after optimizing the structure parameter to realize SHG and THG at the same temperature.

In conclusion, we experimentally demonstrate the coupled conversion of OAM from the THG process in the QPPLT crystal. The QPPLT structure provides the reciprocal vectors to simultaneously fulfill the QPM conditions in the coupled SHG and SFG processes for the efficient generation of a TH OAM beam. Collinear TH and SH beams with different OAMs are obtained in the experiment. Our experimental results prove that the OAM conserves in coupled nonlinear optical conversions. We also find that the conversion efficiency becomes smaller as increasing the topological charge of the input OAM state because of the decreased fundamental power density. The study helps us better understand the OAM conversions in nonlinear optics, which has potential applications in the efficient generation of an OAM state at a short wavelength.

## Funding

This work was supported by National Basic Research Program of China (No. 2012CB921804), National Science Foundation of China (Nos. 11274162, 11274165, and 11321063), and Priority Academic Program Development of Jiangsu Higher Education Institutions (PAPD).

## References

1. L. Allen, M. W. Beijersbergen, R. J. C. Spreeuw, and J. P. Woerdman, Phys. Rev. A **45**, 8185 (1992).
2. L. Marrucci, C. Manzo, and D. Paparo, Phys. Rev. Lett. **96**, 163905 (2006).
3. M. W. Beijersbergen, R. P. C. Coerwinkel, M. Kristensen, and J. P. Wperdman, Opt. Commun. **112**, 321 (1994)
4. N. R. Heckenberg, R. McDuff, C. P. Smith, and A. G. white, Opt. Lett. **17**, 221 (1992)
5. R. K. Tyson, M. Scipioni, and J. Viegas, Appl. Opt. **47**, 6300 (2008).
6. D. G. Grier, Nature 424, 810 (2003).
7. S. Tao, X. C. Yuan, J. Lin, X. Peng, and H. Niu, Opt. Express **13**, 7726 (2005).
8. J. R. Collins, and A. Stuart, J. Opt. Soc. Am. **60**, 1168 (1970).
9. M. P. MacDonald, L. Paterson, K. Volke-Sepulveda, J. Arlt, W. Sibbett, and K. Dholakia, Science **296**, 1101 (2002)
10. S. Barnet, A. Jesacher, S. Furhapter, C. Maurer, and M. Ritsch-Marte, Opt. Express **14**,3792 (2006)
11. A. Mair, A. Vaziri, G. Weihs, and A. Zeilinger, Nature **412**, 313 (2001).
12. A. C. Dada, J. Leach, G. S. Buller, M. J. Padgett, and E. Andersson, Nature Phys. **7**, 677 (2011).
13. J. T. Barreiro, T. C. Wei, and P. G. B. Kwiat, Nature Phys. **4**, 282 (2008).
14. D. S. Ding, Z. Y. Zhou, B. S. Shi, and G. C. Guo, Nat. Commun. **4,** 141 (2013).

15. G. Molina-Terriza, J. P. Torres, and L. Torner, Nature Phys. **3**, 305 (2007).
16. H. Huang, G. Xie, Y. Yan, N. Ahmed, Y. Ren, Y. Yue, D. Rogawski, M. J. Willner, B. I. Erkmen, K. M. Birnhaum, et al. Opt. Lett. **39**, 197 (2014)
17. G. Gariepy, J. Leach, K. T. Kim, T. J. Hammond, E. Frumker, R. W. Boyd, and P. B. Corkum, Phys. Rev. Lett. **113**, 153901 (2014)
18. J. Arlt, K. Dholakia, L. Allen, and M. J. Padgett, Phys. Rev. A **59**, 3950 (1999)
19. S. Feng, and P. Kumar, Phys. Rev. Lett. **101**, 3958 (2008).
20. H. H. Arnaut, and G. A. Barbosa, Phys. Rev. Lett. **85**, 286 (2000).
21. K. Dholakia, N. B. Simpson, M. J. Padgett, and L. Allen, Phys. Rev. A **54**, 3742 (1996).
22. J. Courtial, K. Dholakia, L. Allen, and M. J. Padgett, Phys. Rev. A **56**, 4193 (1997).
23. Z. Y. Zhou, D. S. Ding, Y. K. Jiang, S. Shi, X. S. Wang, B. S. Shi, and G. C. Guo, Opt. Express **22**, 20298 (2014).
24. S. M. Li, L. J. Kong, Z. C. Ren, Y. Li, C. Tu, and H. T. Wang, Phys. Rev. A **88**, 035801 (2013).
25. X. P. Hu, X. Wang, J. L. He, Y. X. Fan, S. N. Zhu, H. T. Wang, Y. Y. Zhu, and N.B.Ming, Appl. Phys. Lett. **85**, 188 (2004).
26. P. Xu, S. H. Ji, S. N. Zhu, X. Q. Yu, J. Sun, H. T. Wang, J. L. He, Y. Y. Zhu, and N. B. Ming, Phys. Rev. Lett. **93**, 133904 (2004)
27. Y. Zhang, Z. D. Gao, Z. Qi, S. N. Zhu, and N. B. Ming, Phys. Rev. Lett. **100**, 163904 2008
28. Y. Zhang, J. M. Wen, S. N. Zhu, and M. Xiao, Phys. Rev. Lett. **104**, 183901 (2010).
29. G. H. Shao, Z. J. Wu, J. H. Chen, F. Xu and Y. Q. Lu, Phys. Rev. A **88**, 063827(2013)
30. Y. Li, Z. Y. Zhou, D. S. Ding, and B. S. Shi, J. Opt. Soc. Am. B **32**, 407 (2015).
31. N. G. R. Broderick, G.W. Ross, H. L. Offerhaus, D. J. Richardson, and D. C. Hanna, Phys. Rev. Lett. **84**, 4345 (2000).
32. X. P. Hu, G. Zhao, C. Zhang, Z. D. Xie, J. L. He, and S. N. Zhu, Appl. Phys. B **87**, 91 (2007)
33. S. N. Zhu, Y. Y. Zhu, and N. B. Ming, Science, **278**, 843 (1997).
34. Y. Q. Qin, Y. Y. Zhu, S. N. Zhu and N. B. Ming, J. Appl. Phys. **84**, 6911 (1998).
35. X. Y. Fang, D. Z. Wei, D. M. Liu, W. H. Zhong, R. Ni, Z. H. Chen, X. P. Hu, Y. Zhang, S. N. Zhu, and M. Xiao, Appl. Phys. Lett. **107**, 161102 (2015).

# Figures

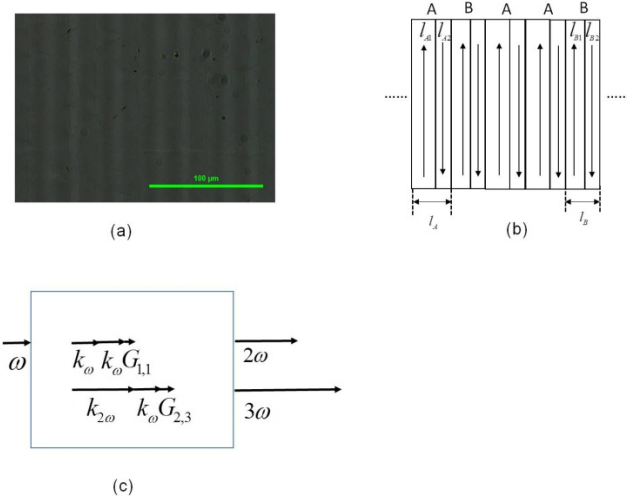

FIG. 1. (a) Microscopic photo of the QPPLT sample. (b) One segment of the structure. (c) Schematic QPM diagram of the THG process in the QPPLT crystal.

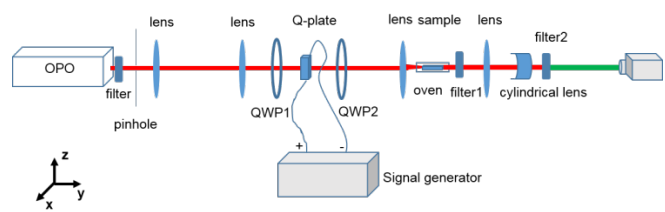

FIG.2. Schematic of the experimental setup.

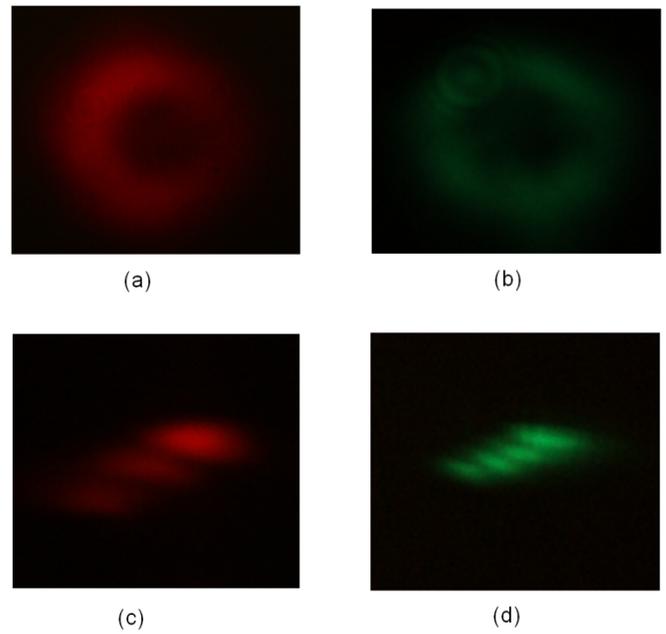

FIG. 3. The SH (a) and TH (b) beams generated by a pump beam carrying an OAM of $l_1 = 1$. By using a cylinder lens, the converted pattern indicates $l_2 = 2$ for the SH beam (c) and $l_3 = 3$ for the TH beam (d).

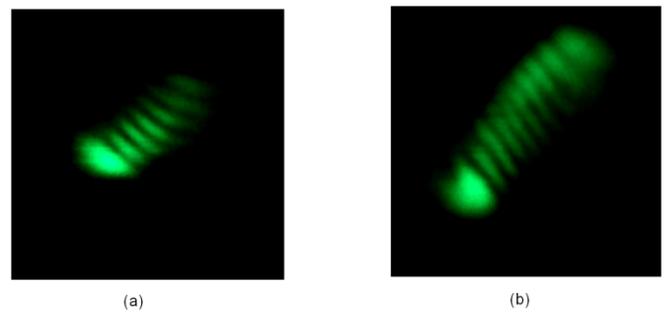

FIG. 4. (a)Interference patterns of THG generated by pump beam with l=2. (b)Interference pattern of THG generated by pump beam with l=3.

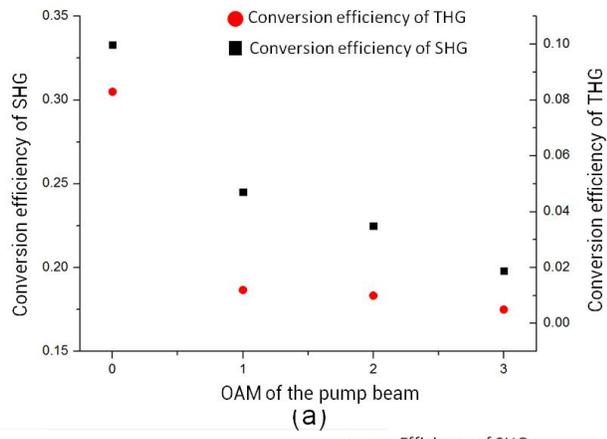

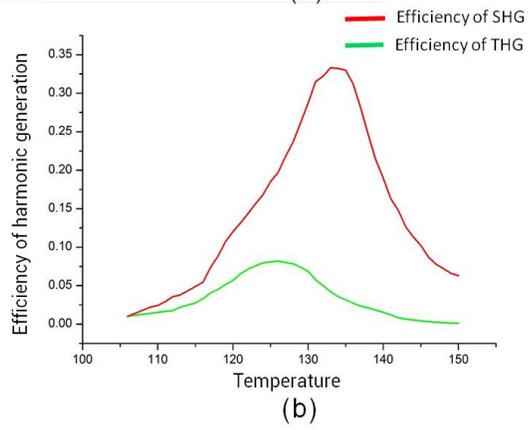

FIG. 5. (a) The conversion efficiencies of SHG and THG pumped by different OAM states. (b) Temperature tuning curves of SHG and THG.